\documentclass[12pt,a4paper]{article}
\usepackage{graphicx}
\begin{document}
\newcommand\bxi {\mbox{\boldmath$\xi$}}
\newcommand\eqref[1]{(\ref{#1})}

{\Large 
Finite Element Method for Solving the Collective Nuclear Model with Tetrahedral Symmetry}
\vspace{5mm}

\centerline{ \large \it 
A.A. Gusev$^{~a}$, S.I. Vinitsky$^{~a,b}$, O.Chuluunbaatar$^{~a,c}$,  A. G\'o\'zd\'z$^{~d}$,
}  \centerline{\large \it 
A. Dobrowolski$^{~d}$, K. Mazurek$^{~e}$, P.M. Krassovitskiy$^{~a,f}$}
\vspace{5mm}

\noindent
{\it  $^{a}$Joint Institute for Nuclear Research, Dubna, Russia\\
$^{b}$RUDN University, 6 Miklukho-Maklaya St., Moscow 117198, Russia\\
$^{c}$Institute of Mathematics, National University of Mongolia, Ulaanbaatar, Mongolia\\
$^{d}$Institute of Physics, University of M. Curie{-}Sklodowska, Lublin, Poland\\
$^{e}$Institute of Nuclear Physics PAN, Krak\'ow, Poland\\
$^{f}$Institute of Nuclear Physics, Almaty, Kazakhstan}
\begin{abstract}
We apply a new calculation scheme of a finite element method (FEM) for for solving an elliptic boundary-value problem
describing a quadrupole vibration collective nuclear model with tetrahedral symmetry.
We use of shape functions constructed with interpolation Lagrange polynomials
on a triangle finite element grid and compare the FEM results with obtained early by a finite difference method.\footnote{Submitted to: Acta Physica Polonica B Proceedings Supplement }
\end{abstract}

\section{Introduction}
In recent papers the consistent approach
to quadrupole{-}octupole collective vibrations coupled with the rotational motion
was presented to find and/or verify some fingerprints of possible high-rank symmetries (e.g., tetrahedral, octahedral, ...) in the recent experimental data of nuclear collective bands
\cite{Dobrowolski2016,Dobrowolski2018}.
A realistic collective Hamiltonian with variable mass{-}parameter tensor and potential obtained through the macroscopic{-}microscopic Strutin\-sky-like method with particle{-}number{-}projected BCS approach in full vibrational and rotational, nine{-}dimensional collective space
was diagonalized in the basis of projected harmonic oscillator eigensolutions.
In this approach the symmetrized orthogonal basis of zero{-}, one{-}, two{-} and three{-}phonon oscillator{-}like functions in vibrational part,
coupled with the corresponding Wigner function \cite{Szulerecka2014} has been applied for solving the boundary value problem (BVP) in 6D domain.
The algorithms for construction the symmetrized basis was considered in \cite{casc15,casc16} w.r.t. symmetrization group
 \cite{Gozdz2011IJMPE,Gozdz2013PhysAtNuc,Gozdz2013PhysScr}.
 In paper \cite{Maz2011} the BVP in 2D domain describing the above quadrupole vibration collective nuclear model of  $^{156}$Dy nucleus
with tetrahedral symmetry \cite{cornwell84}
has been solved by a finite difference method (FDM) that  was a part of the BVP in 6D domain.
However,
the FDM approach did not obtain further generalization on the above multidimensional domain,
where the potential energy and components of the metric tensor given by $2\times 10^{6}$ table values.

In this paper we consider the alternative approach which is applicable for solving the BVP in the multidimensional domain of $d$--dimensional Euclidian space
divided into the $d!$ simplexes in the framework of a finite element method (FEM) with Lagrangian elements and PI-type Gauss quadrature formulas
 in the simplexes \cite{sb,casc17a,casc18}.

An efficiency of the applied finite element calculation scheme is shown by the benchmark calculations of the above BVP in the 2D domain.
We apply shape functions on triangle finite element grid using the interpolation Lagrange polynomials of two variables
with quadrature rules in triangle \cite{casc17b} and compare our
FEM results with obtained early by the FDM \cite{Maz2011}.
\begin{figure}[t]
\includegraphics[width=0.31\textwidth]{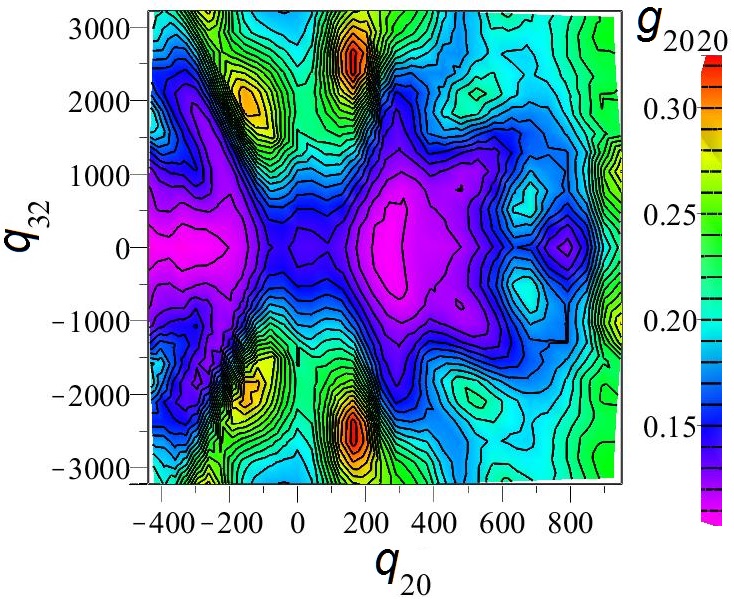}a
\includegraphics[width=0.31\textwidth]{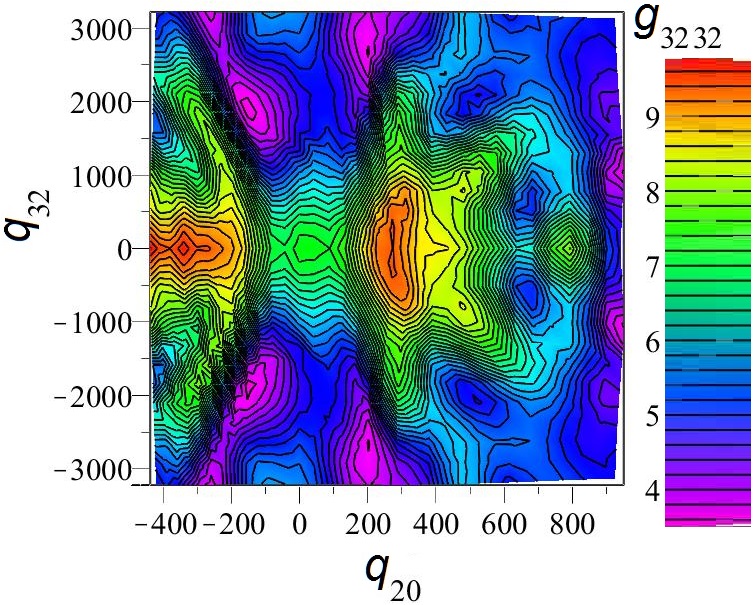}b
\includegraphics[width=0.31\textwidth]{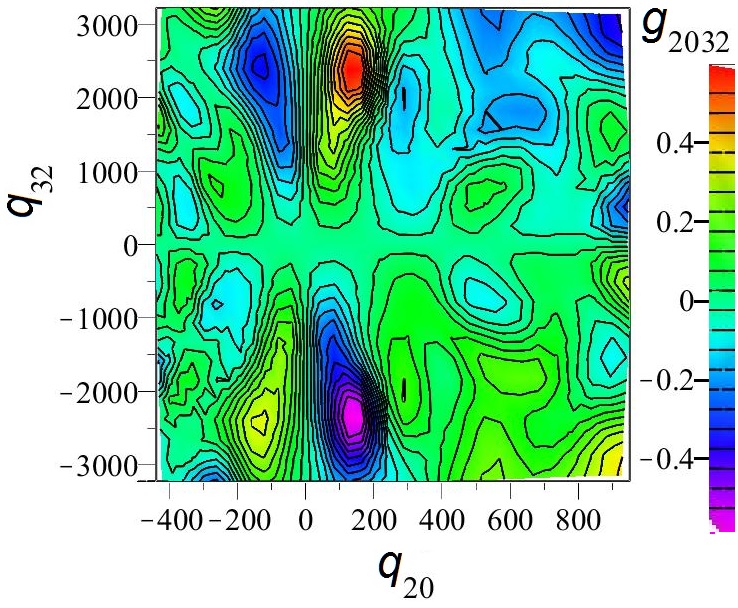}c\\
\includegraphics[width=0.33\textwidth]{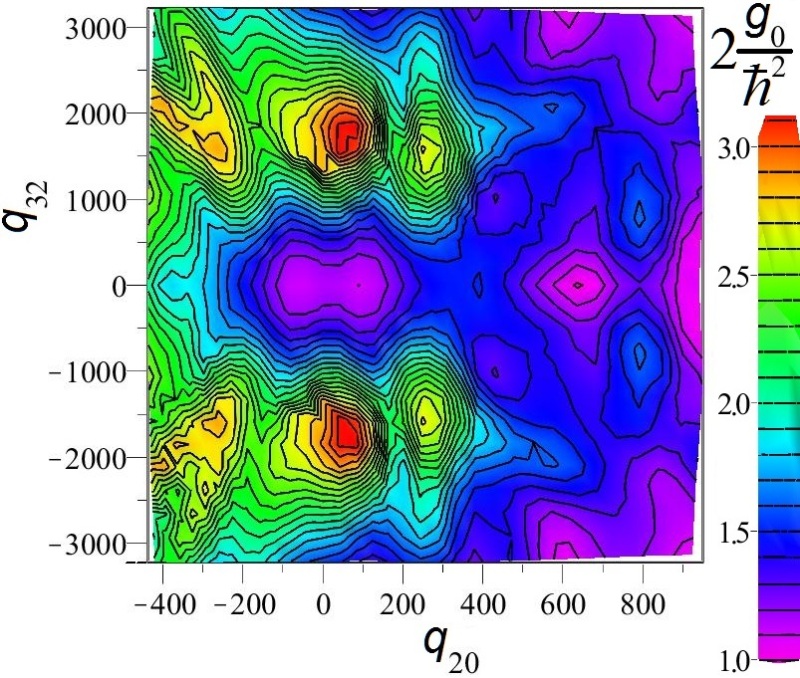}d
\includegraphics[width=0.48\textwidth]{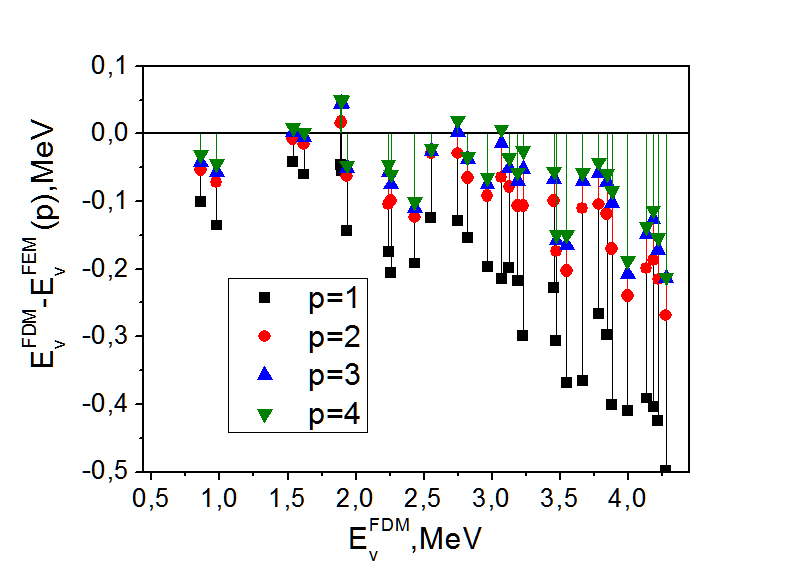}e
\caption{
The coefficients $g_{ij}(x)$ from (\ref{43}) given  in variables $(q_{20},q_{32})$ (a,b,c).
Square root of the determinant $2g_{0}(x)/\hbar^2{=}\sqrt{{\rm det}B(q_{20},q_{32})}$   constructed out of collective inertia parameters in units $10^{-5}\hbar^2/$(MeV fm$^5$)(d).
The differences $E_v^{\rm FDM}-E_v^{\rm FEM}(p)$
between
eigenvalues of $E_v^{\rm FDM}$ of $^{156}$Dy nucleus calculated by the FDM
\cite{Maz2011} and $E_v^{\rm FEM}(p)$
calculated in the present paper by FEM with triangular Lagrange elements of the order $p=1,2,3,4$  for 30 lowest states
of the  BVP (\ref{1n})--(\ref{43}) in variables $(q_{20},q_{32})$ (e).
}\label{fignya2}
\end{figure}


\section{The setting of the problem}

Consider a self-adjoint boundary-value problem for the elliptic differential equation of the second order \cite{sb,casc18}:
 \begin{eqnarray}
   ( {D}-
     E ) \Phi(x)\equiv
   \bigg(-  {\frac{1}{g_0(x)}}\sum\nolimits_{ij=1}^d\frac{\partial}{\partial x_i}g_{ij}(x)
     \frac{\partial }{\partial x_j}
     +  {V}(x)
     -E \bigg)\Phi(x)=0.\label{1n}
\end{eqnarray}
It is also assumed that $g_0(x)>0$, $g_{ji}(x)=g_{ij}(x)$  and ${V}(x)$ are real-valued functions, continuous together with their generalized derivatives
to a given order in the domain
$x\in\bar\Omega=\Omega\cup\partial\Omega$ with the piecewise continuous boundary $S=\partial\Omega$,
which provides the existence of nontrivial solutions obeying the boundary conditions
of the first kind (I) {or the second  kind (II)}:
\begin{eqnarray}
\label{2n}
(I)~  \Phi(x)\Bigl|_{S}{=}0,
  \label{2a}
~  (II)~  \frac{\partial\Phi(x)}{\partial n_{D}}\Bigl|_{S}{=}0,
\quad
    \frac{\partial\Phi(x)}{\partial
    n_{D}}{=}\sum\nolimits_{ij=1}^d(\hat n, \hat e_i)
  g_{ij}(x)\frac{\partial\Phi(x)}{\partial x_j}.
\end{eqnarray}
Here $\frac{\partial\Phi_m(x)}{\partial n_{D}}$ is the derivative along the
conormal direction, $\hat n$ is the outer normal to the boundary
 of the domain $S=\partial\Omega$, $\hat e_i$ is the unit vector of
$x=\sum\nolimits_{i=1}^d \hat e_i x_i$, and $(\hat n, \hat e_i)$ is the scalar product in ${\cal R}^d$.
It is also assumed that the metric tensor $g_{ij}(x)$ is positively defined what implies the positive  determinant $\mathrm{det}(g_{ij}(x)) >0$.

For a discrete spectrum problem, the functions $\Phi_m(x)$ from the Sobolev space $H_2^{s\geq1}(\Omega)$,
$\Phi_m(x)\in H_2^{s\geq1}(\Omega)$, corresponding to the real eigenvalues $E$: $E_1\leq E_2\leq\ldots\leq E_m\leq\ldots $
satisfy the conditions of normalization and orthogonality
 \begin{eqnarray}\label{3}
   \langle \Phi_m(x)| \Phi_{m'}(x)\rangle
   =\int\nolimits_{\Omega}dx g_0(x){\Phi}_m(x)\Phi_{m'}(x)=\delta_{mm'},
   \quad dx=dx_1\ldots dx_d.
\end{eqnarray}

The FEM solution of the boundary-value problems (\ref{1n})--(\ref{3})
is reduced to the determination of stationary points of the variational functional
\cite{sb,casc18}
 \begin{eqnarray}
\label{4}
  \Xi (\Phi_m,E_m) \equiv\int\nolimits_{\Omega} dx g_0(x){\Phi}_m(x)
  \left( {D}-E_m \right)\Phi(x)
   =\Pi(\Phi_m,E_m),
 \end{eqnarray}
where
$\Pi(\Phi,E)$ is the symmetric quadratic functional
 \begin{eqnarray*}
\label{symmr}
\Pi(\Phi,E)= \int\nolimits_{\Omega}dx\biggl[
\sum\nolimits_{ij=1}^dg_{ij}(x)\frac{\partial {\Phi}(x)}{\partial
  x_i} \frac{\partial \Phi (x)}{\partial x_j}  +
g_0(x){\Phi}(x) ({V}(x)
-E ) \Phi (x) \biggr].
\end{eqnarray*}

\begin{figure}[t]
\includegraphics[width=0.44\textwidth]{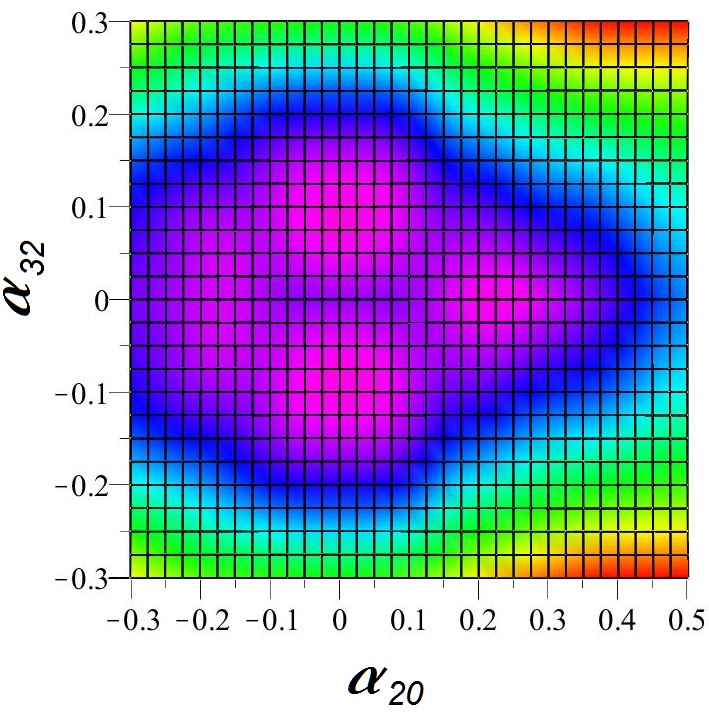}a \includegraphics[width=0.52\textwidth]{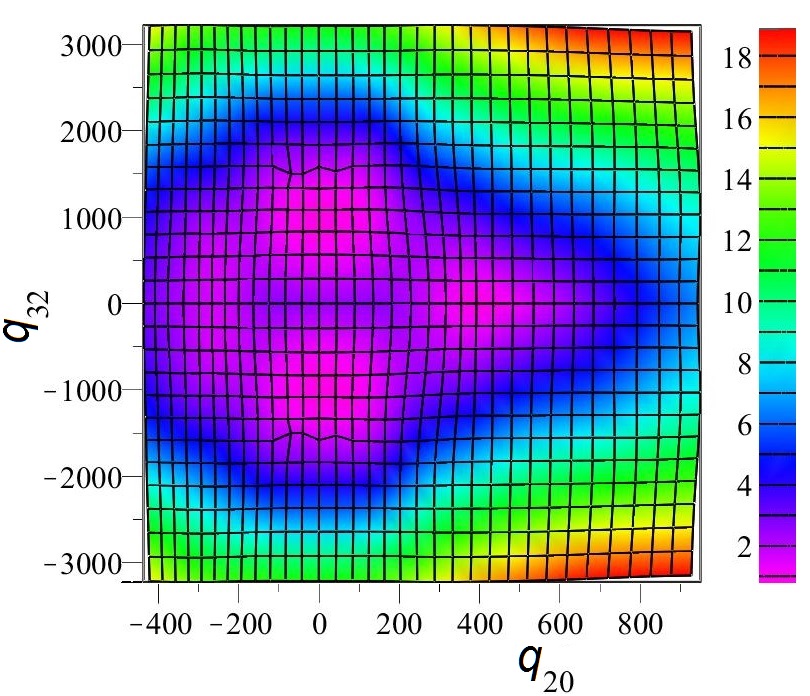}b
\caption{The potential energy $V(x_1,x_2)$
of $^{156}$Dy nucleus
given in variables $(\alpha_{20},\alpha_{32})$ (a) and in variables $(q_{20},q_{32})$ (b).
The nodal points of finite element grid are intersection points of horizontal and vertical lines.}\label{fignya1}
\end{figure}

\section{Quadrupole-octupole-vibrational collective model}
Below we solve the  BVP (\ref{1n})--(\ref{3}) in the 2D domain $d=2$ that describe the quadrupole-octupole-vibrational collective model
of  $^{156}$Dy nucleus \cite{Maz2011} with the coefficients
$g_{0}(x)$ and
$g_{ij}(x)$ determined  by the expressions $i,j=1,2$:
 \begin{eqnarray}
g_{0}(x_1,x_2){=}\frac2{\hbar^2}\sqrt{{\rm det}B(x_1,x_2)},
~
g_{ij}(x_1,x_2){=}\sqrt{{\rm det}B(x_1,x_2)}[B^{-1}(x_1,x_2)]_{ij}.\label{43}
\end{eqnarray}
The mass tensor  $B_{ij}(x_1,x_2)$ has been calculated \cite{Maz2011} in the terms of the average nuclear deformations $x=(x_1,x_2)=(q_{20},q_{32})$
determine in \cite{AGB}, and shown in Fig. \ref{fignya2}a-d.
The potential energy function $V(x_1,x_2)$   has been calculated  in the terms of the   nuclear deformations
 $x=(x_1,x_2)=(\alpha_{20},\alpha_{32})$ \cite{Maz2011}
and shown  in these coordinates as well as in coordinates $x=(x_1,x_2)=(q_{20},q_{32})$ in Fig. \ref{fignya1}a,b.

\begin{table}
\caption{The low part of the spectrum of 10 lowest states of
$^{156}$Dy nucleus
counted from minimum of potential energy ($V_{\rm min}(\alpha_{20},\alpha_{32})=0.685 MeV$).
$E_v^{\rm FDM}$ calculated by FDM of the second order \cite{Maz2011} and $E_v^{\rm FEM}(p)$
calculated by FEM
with triangular Lagrange elements of the order $p=1,2,3,4$ in the present paper.}\label{tabl1}

\begin{tabular}{|rrrrrr|} \hline
$v$&$E_v^{\rm FDM}$& $E_v^{\rm FEM}(1)$&$E_v^{\rm FEM}(2)$&$E_v^{\rm FEM}(3)$&$E_v^{\rm FEM}(4)$\\  \hline
1&0.85988&0.96000&0.91329&0.90234& 0.89065\\ 
2&0.97588&1.11144&1.04808&1.03297& 1.02068\\ 
3&1.53669&1.57813&1.54403&1.53371& 1.52776\\ 
4&1.61774&1.67776&1.63332&1.62287& 1.61571\\ 
5&1.88907&1.93560&1.87335&1.84504& 1.83794\\ 
6&1.89469&1.94932&1.87706&1.84925& 1.84631\\ 
7&1.93369&2.07731&1.99714&1.98486& 1.98032\\ 
8&2.23907&2.41405&2.34335&2.29594& 2.28444\\ 
9&2.25778&2.46383&2.35681&2.33287& 2.31778\\ 
10&2.43288&2.62454&2.55679&2.54278& 2.53388\\ 
 \hline
 \end{tabular}
\end{table}
\begin{figure}[t]
\centerline{1\phantom{0000000000000}2\phantom{0000000000000}3\phantom{0000000000000}4\phantom{0000000000000}5}
\centerline{\includegraphics[width=0.19\textwidth]{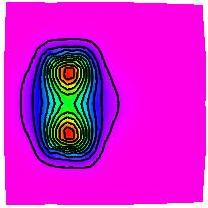}
\includegraphics[width=0.19\textwidth]{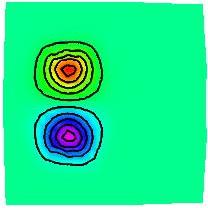}
\includegraphics[width=0.19\textwidth]{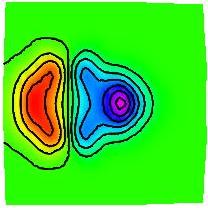}
\includegraphics[width=0.19\textwidth]{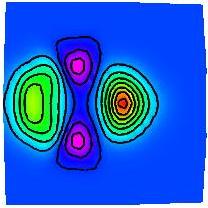}
\includegraphics[width=0.19\textwidth]{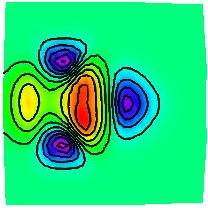}}
\centerline{6\phantom{0000000000000}7\phantom{0000000000000}8\phantom{0000000000000}9\phantom{000000000000}10}
\centerline{\includegraphics[width=0.19\textwidth]{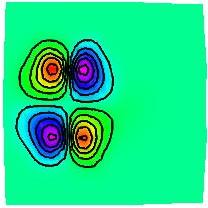}
\includegraphics[width=0.19\textwidth]{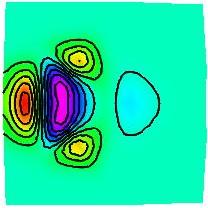}
\includegraphics[width=0.19\textwidth]{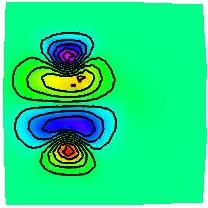}
\includegraphics[width=0.19\textwidth]{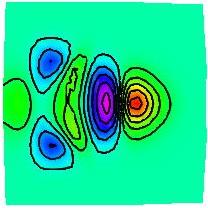}
\includegraphics[width=0.19\textwidth]{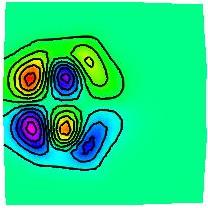}}
\caption{The first ten eigenfunctions of $^{156}$Dy nucleus in the plane $(q_{20},q_{32})$}\label{fignya3}
\end{figure}
Table \ref{tabl1} shows a low part of the spectrum of $v=1,...,10$ states of  $^{156}$Dy
counted from minimum of potential energy ($V_{\rm min}(\alpha_{20},\alpha_{32})=0.685 MeV$).
Second column shows eigenenergies $E_v^{\rm FDM}$ calculated by the FDM code of the second order \cite{Maz2011}.
The remaining columns show the eigenvalues $E_v^{\rm FEM}(p)$ of the BVP (\ref{1n})-(\ref{43})
in $\Omega(q_{20},q_{32})$ with coefficients $g_{ij}(q_{20},q_{32})$
determined by formulas (\ref{43})
and the potential energy functions $V(q_{20},q_{32})$ calculated in the present paper
by the FEM code with
the Gaussian quadratures PI type till the eight order \cite{casc17b}.
Calculations has been carried out with the second type (II) boundary conditions (\ref{2a})
and orthonormalization condition (\ref{3}) with triangular Lagrange elements of the order $p=1,2,3,4$ in the finite-element grid $\Omega(q_{20},q_{32})$.
Discrepancy $E_v^{\rm FDM}-E_v^{\rm FEM}(p)$ between
the results of FDM and FEM calculations in dependence of the order $p=1,2,3,4$ of the FEM approximation
is shown in Fig. \ref{fignya2}e.
One can see that in increasing the order of the FEM approximation the discrepancy is decreased till 1\%.
Fig. \ref{fignya3} display the corresponding eigenfunctions
 $\Phi_v(q_{20},q_{32})$
 in the finite-element grid $\Omega(q_{20},q_{32})$.
 The eigenfunctions of
 the ground and first excited states are in   good agreement  with the eigenfunctions calculated
 in domain $\Omega(\alpha_{20},\alpha_{32})$ by the FDM
 \cite{Maz2011}.
The third eigenfunction
has one node line in direction $\alpha_{20}$ in contrast with the third  FDM eigenfunction that has no nodes.
Meanwhile, the forth function has two node lines in direction of $\alpha_{20}$ and qualitative coincides with the forth FDM eigenfunction.
We can suppose that the revivable  distinctions  are consequence of approximation of table values of $V(\alpha_{20},\alpha_{32})$  on the FEM grid $\Omega(q_{20},q_{32})$ instead of approximation of derivatives of table values of $g_{ij}(q_{20},q_{32})$ on the FDM grid $\Omega(\alpha_{20},\alpha_{32})$ accepted in \cite{Maz2011}.
\section{Conclusion}
We applied the new calculation schemes in the framework of FEM with the triangular Lagrange elements and Gaussian quadratures for analysis of the quadrupole vibration collective nuclear model with tetrahedral symmetry.
We constructed of shape functions on triangle finite element grid and compared
our FEM results with obtained early by FDM that are in a good agreement.
This approach is generalized directly for the solving  BVP
in multidimensional domain by using the algorithms and their program realization \cite{casc17a,casc18}.
We will apply the proposed FEM  for solving the BVP in the six dimensional domain describing the above quadrupole{-}octupole collective vibration model,
 in our further papers.


This work was supported by the Polish--French COPIN
collaboration of the project 04--113,
Bogoliubov--Infeld and Hulubei--Meshcheryakov JINR programs,
the grant RFBR 18--51--18005, RUDN University Program 5--100 and
grant of Plenipotentiary of the Republic of Kazakhstan in JINR.

\end{document}